\documentclass [12pt]{article}
\def\maj#1{\ifmmode\mbox{\usefont{U}{msb}{m}{n}#1}\else{\usefont{U}{msb}{m}{n}#1}\fi}
\def\v#1{\mathbf{#1}}

\begin{document}

\title{\textbf{Major difference\\ between true bosons and
``proteons''}}
\author{M. Combescot and O.
Betbeder-Matibet\\
\small{\textit{GPS, Universit\'e
Pierre et Marie Curie and Universit\'e Denis
Diderot, CNRS,}}\\
\small{\textit{Campus Boucicaut, 140 rue de
Lourmel, 75015 Paris, France}}}
\date{}
\maketitle

\begin{abstract}
We call ``proteons'' --- from the ever-changing greek sea-god
$\Pi\rho\omega\tau\epsilon\upsilon\varsigma$ --- composite
particles made of two fermions. Among them, are the
semiconductor excitons, but also various atoms and molecules,
like the giant molecules made of two
$^{40}$K or
$^6$Li atoms which have recently Bose condensed. In
addition to their indistinguishability, these composite
particles are ``ever-changing'' in the sense that there is no
way to know with which fermions they are precisely made. As
direct consequences, (i) the proteons are not true bosons,
(ii) the basis made with proteon states is \emph{overcomplete}.
In spite of these difficulties, these proteons do have a nice
closure relation, unexpected at first, \emph{different from the
boson one} and which makes the bosonization procedures used up
to now to treat many-body effects between composite bosons,
rather questionable, due to possibly incorrect sum rules 
resulting from it. This closure relation in particular
explains, in a neat way, the surprising factor
1/2 between the inverse lifetime
and the sum of scattering rates which exists for exact 
excitons but not for boson excitons, as we have recently shown. 
\end{abstract}

\vspace{2cm}

PACS.: 71.35.-y Excitons and related phenomena

\newpage

The excitons, \emph{i}.\ \emph{e}., the \emph{one}
electron-hole (e-h) pair eigenstates of the semiconductor
Hamiltonian, are usually considered as bosons. And indeed,
various attempts have been devoted to the
experimental observation [1-3] of their Bose-Einstein
condensation. However, if we are interested in many-body
effects, we must be aware that excitons are composite
particles, so that they feel each other, not only through
Coulomb interactions between carriers, but also through Pauli
exclusion between the fermions which compose them. This Pauli
way for excitons to interact, which does not need any Coulomb
process to exist, is $N$-body ``at once'': Indeed, any new
exciton feels \emph{all} the excitons present in the
sample since it has to be made with  fermion states different
from the ones already used to make the previous excitons. This
``Pauli interaction'', which is a direct consequence of the
fact that the excitons are not true bosons, induces effects
as large as the ones induced by Coulomb
interactions. They can even be dominant, as we have recently
shown [4], in the case of the semiconductor nonlinear
susceptibility $\chi^{(3)}$.

In order to take exactly into account this Pauli way for
excitons to interact, we have recently developed a new many-body
theory for composite bosons [5-8]. It relies on two
scatterings: $\xi_{mnij}^\mathrm{dir}$ corresponds to
\emph{direct Coulomb} processes between two excitons, the
``in'' and ``out'' excitons $(i,j)$ and $(m,n)$ being made with
the same pairs, while $\lambda_{mnij}$ corresponds to a
\emph{fermion exchange} between the ``in'' and ``out''
excitons, \emph{in the absence of any Coulomb process}.

This new procedure has to be contrasted with the previous
approaches in which the exact semiconductor Hamiltonian $H$
is replaced by an effective Hamiltonian $H_\mathrm{eff}$
written in terms of boson-exciton operators, the composite
nature of the original excitons being hidden in the two-body
potential of this Hamiltonian, through a scattering
``dressed by exchange''. Even if the $H_\mathrm{eff}$ widely
used [9] by semiconductor physicists has to be rejected because
it is not hermitian --- a major failure apparently missed by
every one --- the concept of an effective bosonic
Hamiltonian for excitons is very appealing at first: Indeed, if
we manage to replace
$H$ by
$H_\mathrm{x}+V_\mathrm{xx}$, we can then treat
many-body effects between excitons in a standard way, as all
known procedures dealing with interactions rely on a
perturbative expansion in the interacting potential.
Unfortunately, such a $V_{XX}$ potential does not exist for
composite excitons: Indeed, $V_{eh'}$ has to be considered as a
part of a potential between two excitons if we see them as 
made of
$(e,h)$ and
$(e',h')$, while the same $V_{eh'}$ is an ``internal'' potential
if the  excitons are made of
$(e,h')$ and $(e',h)$. Since the carriers are
indistinguishable, there is no way to know !

We have first questioned the concept of
bosonized Hamiltonian when we calculated correlation
effects between excitons [7]. The major alert however came very
recently [10], when we compared the inverse lifetime and the sum
of scattering rates of exciton states, calculated with exact
excitons using our new many-body theory, and with bosonized
excitons using an effective bosonic Hamiltonian: An
unexpected additional factor 1/2 appears for exact excitons,
which makes any agreement between the exact results and the ones
obtained with boson-excitons, impossible,
\emph{whatever the scatterings used in  $V_\mathrm{xx}$ are}.

When we bosonize the excitons, we irretrievably lose their
composite nature. The purpose of this letter is to show that
it is not enough to repair this loss by the introduction
of appropriate X-X scatterings ``dressed by exchange''. The
composite nature of the excitons is quite deep. It also
appears through the closure relation between exciton states,
which is definitely different from the one of boson-excitons.
At first, we did not expect such a closure
relation for exact exciton states to exist, because these
states are somewhat ticklish : (i) They are non-orthogonal;
(ii) worse, they form an overcomplete set. Indeed, for
$N=2$ already, we have (see eq.\ (7) of ref.\ [6])
\begin{equation}
\langle v|B_mB_nB_i^\dag
B_j^\dag|v\rangle=\delta_{mi}\delta_{nj}+\delta_{mj}
\delta_{ni}-2\lambda_{mnij}\ ,
\end{equation}
where $B_i^\dag$ is the creation operator of the exact exciton
$i$, (orthogonal states would only have the $\delta$ terms),
while the overcompleteness of these exciton states follows from
\begin{equation}
B_i^\dag B_j^\dag=-\sum_{mn}\lambda_{mnij}\,B_m^\dag B_n^\dag
\ ,
\end{equation}
(see eq.\ (5) of ref.\ [6]) which results from the two ways
to make two excitons out of two e-h pairs. This overcompleteness
is actually trivial to see: If
$\mathcal{N}$ is the number of electron (or hole) states in the
sample, the possible number of e-h pairs is $\mathcal{N}^2$,
just as the possible number of $i$ excitons, since both, the
free pairs and the excitons, are eigenstates of the same
one-pair Hamiltonian. If we now  turn to two-free-pair states,
their number is
$\mathcal{N}^2(\mathcal{N}-1)^2$, due to Pauli
exclusion, while the possible number of two-exciton
states $(i,j)$ is $\mathcal{N}^4$: For large $\mathcal{N}$,
there are essentially $2\mathcal{N}^3$ exciton states in excess
!

It is clear that,
if a closure relation, different from the
boson-exciton one, has to exist for exact-exciton states, to
bosonize the excitons must have dramatic consequences: Indeed,
all sum rules which rely on this closure relation, will be
different for exact and boson excitons, making the
agreement of the corresponding quantities impossible.

In this letter, we are going to speak in terms of excitons,
because they are the familiar bosons for us. In addition,
their Hamiltonian is nicely simple, being the one of fermions
in Coulomb interaction, so that they are under complete
control. We however wish to stress that the question raised
here, has a much wider interest. Indeed, particles
composed of fermions which look like bosons, exist in many
other fieds than semiconductor physics, and attempts to
bosonize them always have had the same purpose: 
to possibly handle their interactions, in the absence of a
many-body theory which can take care of their composite nature
properly. Among composite bosons of current interest, we can
cite the molecules made of two
$^{40}$K or $^6$Li atoms for which a Bose-Einstein
condensation has recently been observed [11-14]. Compared to
$^7$Li atoms for which a BEC was first observed [15-16], these
molecules are quite large so that their interactions, in
particular through ``Pauli scatterings'', are expected to
play a role as important as for excitons.

In order to give a wider audience to what is here said on
excitons, it thus appeared to us of interest to introduce a new
particle concept: \emph{the proteon}. Like the ever-changing
greek sea-god $\Pi\rho\omega\tau\epsilon\upsilon\varsigma$
(Proteus), there is no way to know with which particular
fermions these particles are made. Being moreover
indistinguishable, these proteons in fact correspond to a new
type of quantum particles, far more tricky than the fermions or
the (true) bosons: Indeed, due to their ever-changing
character, there is no way to write down an
\emph{interacting potential between them}, so that completely
new procedures have to be constructed in order to treat their
many-body effects properly. This is the goal of the set of
works we are presently doing.

\vspace{0.5cm}

\noindent \emph{Boson-excitons versus exact-excitons}

If the excitons were true bosons, their creation operators,
$\bar{B}_i^\dag$, would obey the commutation rule
$[\bar{B}_j,\bar{B}_i^\dag]=\delta_{ij}$. From it, we can show 
that the $N$-boson-exciton states $\bar{B}_{i_1}^\dag\cdots
\bar{B}_{i_N}^\dag|v\rangle$ form an \emph{orthogonal} basis,
with a closure relation given by
\begin{equation}
I=\frac{1}{N!}\sum_{i_1\cdots i_N}\bar{B}_{i_1}^\dag\cdots
\bar{B}_{i_N}^\dag|v\rangle\langle v|\bar{B}_{i_N}\cdots
\bar{B}_{i_1}\ ,
\end{equation}
as easy to check by inserting this identity into 
$\langle v|\bar{B}_{j_N}\cdots\bar{B}_{jÑ1}|\bar{\psi}
\rangle$, where $|\bar{\psi}\rangle$ is an arbitrary
$N$-boson state. Using eq.\ (3), we can then
expand any $|\bar{\psi}\rangle$ on the $N$-boson-exciton
states in an unique way, due to the orthogonality of these
$\bar{B}_{i_1}^\dag\cdots
\bar{B}_{i_N}^\dag|v\rangle$ states.

For exact-excitons, the situation is far more subtle: 

(i) Using
\begin{equation}
[D_{mi},B_j^\dag]=2\sum_n\lambda_{mnij}\,B_n^\dag\ ,
\end{equation}
where $D_{mi}=\delta_{mi}-[B_m,B_i^\dag]$ is the
``deviation-from-boson operator'', it is formally possible to
calculate the scalar product of 
$N$-exciton states $B_{i_1}^\dag\cdots B_{i_N}^\dag|v\rangle$:
For $N=2$, it is given by eq.\ (1) --- which immediately
follows from eq.\ (4). For large $N$, these scalar products are
far more tricky. When all the excitons are in the same
ground state 0, we have already shown [5] that
$\langle v|B_0^NB_0^{\dag N}|v\rangle$ reads $N!F_N$. While
$F_N$ reduces to 1 for boson-excitons, for exact excitons and
$N$ large, it is a complicated function of $N$ and
$\eta=Na_X^3/\mathcal{V}$, which, in the small $\eta$
limit, behaves [17] as $\exp -(33\pi N\eta/4)$ in 3D. Although
$\eta$ has to be small for the excitons to
exist, $N\eta$ can be much larger than 1 for large samples,
making $F_N$ extremely small. In previous works, we
have also calculated the scalar products of $N$-exciton
states with one or two excitons possibly outside the ground
state [18,19]. Their expressions do not reduce to products of
Kronecker symbols as for orthogonal states, but read in terms of
$N$,
$F_{N-p}$ and the $\lambda_{mnij}$'s in a very complicated way.

(ii) In addition, the decomposition of a
$N$-electron-hole-pair state on these $N$-exciton states is
\emph{a priori} not unique: Due to eq.\ (2), we 
for example have,
\begin{equation}
B_0^{\dag N}
|v\rangle=-\sum_{mn}\lambda_{mn00}B_{m}^\dag
B_{n}^\dag B_0^{\dag N-2}|v\rangle\ .
\end{equation}

In spite of these obvious difficulties, the
$N$-exciton states are physically
appealing in the low density limit
because, in this limit, the e-h pairs are
known to form excitons. If we
for example consider the
$N$-pair ground state, its representation in terms of free
electrons and free holes has to contain
a lot of e-h states with
essentially equal weight, while its representation in terms of
excitons should be mainly made of $B_0^{\dag N}|v\rangle$.

However, in order for these exciton states to be of
practical use, we must find a
simple procedure to express any
$N$-pair state $|\psi\rangle$ on
these $N$-exciton states. Although the
idea was far from our mind at first, a
closure relation, as simple as eq.\
(3), in fact exists for
exact-excitons.  

\vspace{0.5cm}

\noindent \emph{Closure relation for exciton states}

Let us first consider a two-e-h--pair
state $|\psi\rangle$. Its expansion on two-exciton states
must read 
$|\psi\rangle=\sum_{i,j}\psi_{ij}B_i^\dag
B_j^\dag|v\rangle$. The $\psi_{ij}$'s 
\emph{a priori} depend on 
$|\psi\rangle$ and on the indices $(i,j)$
of the $B_i^\dag B_j^\dag|v\rangle$ state
of interest in the expansion. The
simplest-minded idea is to try
$\psi_{ij}=\langle v|B_iB_j|\psi\rangle$
with possibly an additional $(i,j)$ dependent
prefactor. At first, we could think of a prefactor being a
constant
$\alpha$ divided by $\langle v|B_iB_jB_i^\dag
B_j^\dag|v\rangle$, in order for the two-exciton
state $B_i^\dag B_j^\dag|v\rangle$ to appear in a normalized
form in the expansion. It
turns out that this is not correct: The
prefactor of $\langle
v|B_iB_j|\psi\rangle$ is (1/4) for all
$(i,j)$, the correct expansion being
\begin{equation}
|\psi\rangle=\frac{1}{4}\sum_{ij}B_i^\dag B_j^\dag|v\rangle
\langle v|B_iB_j|\psi\rangle\ ,
\end{equation}                  
as easy  to check by calculating the scalar product of
this $|\psi\rangle$ with $\langle v|B_mB_n$, using
eqs.\ (1,2).
From the equivalent of eq.\ (1) for three-exciton states, it is
possible to show that the expansion of three-pair
states is quite similar to eq.\ (6), with a prefactor (1/36)
instead of (1/4). 

This led us to think that the closure relation
for
$N$-exciton states should read
\begin{equation}
I=\frac{1}{(N!)^2}\sum_{i_1\cdots i_N}B_{i_1}^\dag\cdots
B_{i_N}^\dag |v\rangle\langle v|B_{i_N}\cdots B_{i_1}\ .
\end{equation}
In order to prove it, the procedure used for $N=(2,3)$ is
however inappropriate because we do not know how to write the
scalar product of $N$-exciton states in a compact form.

Another way to show it, is to expand the
exciton operators in terms of electrons and holes,
\begin{equation}
B_i^\dag=\sum_{\v k_e,\v k_h}\langle\v k_e,\v k_h|\phi_i
\rangle\,a_{\v k_e}^\dag b_{\v k_h}^\dag\ ,
\end{equation}
where $|\phi_i\rangle$ is the $i$ exciton wave function, and
to use this expansion in the r.h.s.\ of eq.\ (7). This leads to
\begin{eqnarray}
\frac{1}{(N!)^2}\sum_{\{i_n\}}B_{i_1}^\dag\cdots B_{i_N}^\dag
|v\rangle\langle v|B_{i_N}\cdots B_{i_1}=\frac{1}{(N!)^2}
\sum_{\{\v k_n\},\{\v k_n'\},\{\v p_n\},\{\v p_n'\}}
a_{\v k_1}^\dag b_{\v k_1'}^\dag\cdots a_{\v k_N}^\dag
b_{\v k_N'}^\dag|v\rangle\nonumber\\
\langle v|b_{\v p_N'} a_{\v p_N}\cdots b_{\v p_1'} a_{\v p_1}
\prod_{n=1}^N\left(\sum_{i_n}\langle\v k_n,\v
k_n'|\phi_{i_n}\rangle
\langle\phi_{i_n}|\v p_n',\v p_n\rangle\right)\ .
\end{eqnarray}
The summation over $i_n$ can be done through the closure
relation which exists between \emph{one}-exciton states. It
gives
$\delta_{\v k_n,\v p_n}
\delta_{\v k_n',\v p_n'}$, so that the r.h.s.\  of eq.\ (9)
reduces to
\begin{equation}
\frac{1}{(N!)^2}\sum_{\{\v k_n\},\{\v k_n'\}}a_{\v k_1}^\dag
b_{\v k_1'}^\dag\cdots a_{\v k_N}^\dag b_{\v k_N'}^\dag|v
\rangle\langle v|b_{\v k_N'} a_{\v k_N}\cdots b_{\v k_1'}
a_{\v k_1}\ ,
\end{equation}
which is nothing but $I$, due to the closure relation which
exists between electron states, namely
\begin{equation}
I=\frac{1}{N!}\sum_{\v k_n}a_{\v k_1}^\dag\cdots a_{\v k_N}
^\dag|v\rangle\langle v|a_{\v k_N}\cdots a_{\v k_1}\ ,
\end{equation}
and the similar one between hole states. 

This derivation
shows in a transparent way that the prefactor $(1/N!)^2$,
instead of $(1/N!)$, found in the closure
relation of exact excitons is nothing but the direct signature
of the composite nature of these excitons. When they are
bosonized, this composite nature is lost by construction and
the closure relation for $N$ boson-excitons  has the usual
$(1/N!)$ prefactor, characteristic of elementary quantum
particles.

\vspace{0.5cm}

\noindent \emph{Link between the lifetime and the sum of
scattering rates of excitons}

In the light of these different closure relations, let us
reconsider the link between the lifetime and the sum
of scattering rates of exact and boson excitons we recently
found [10]. In order to point out the physical origin of the
striking factor 1/2, which appears for exact excitons, in the
easiest way, we will consider $N=2$ excitons only.

The closure relation for two exact excitons, given in eq.\ (6),
also reads  
\begin{equation}
I=\frac{1}{4}\left[\sum_i(2-2\lambda_{iiii})|\phi_{ii}\rangle
\langle\phi_{ii}|+\sum_{i\neq j}(1-2\lambda_{ijij})|\phi_{ij}
\rangle\langle\phi_{ij}|\right]\ ,
\end{equation}
where, according to eq.\ (1), the $|\phi_{ij}\rangle$'s,
defined as $|\phi_{ij}\rangle=(1+\delta_{ij}-2\lambda_{ijij})^{-1/2}
B_i^\dag B_j^\dag|v\rangle$, are the normalized two-exciton
states.

As the standard Fermi golden rule cannot be used to obtain the
lifetime and scattering rates of exact excitons, because there
is no way to write down an interacting potential \emph{between}
excitons, we have been forced to generate unconventional
expressions of these quantities in which only enters the
Hamiltonian $H$. Of course, these expressions can 
also be used for boson excitons. This led us to
look for the time evolution of an initial state
$|\psi_{t=0}\rangle$ as
\begin{equation}
|\psi_t\rangle=e^{-i(H-\langle H\rangle)t}|\psi_{t=0}\rangle
=|\psi_{t=0}\rangle+|\tilde{\psi}_t\rangle\ ,
\end{equation}
where $\langle H\rangle
=\langle\psi_{t=0}|H|\psi_{t=0}\rangle$ just adds an irrelevant
phase factor to the standard expression of $|\psi_t\rangle$.
For $|\psi_{t=0}\rangle=|\phi_{00}\rangle$, the state
$|\tilde{\psi}_t\rangle$, precisely given by 
\begin{equation}
|\tilde{\psi}_t\rangle=F_t(H-\langle H\rangle)[H-\langle
H\rangle]|\phi_{00}\rangle\ ,
\end{equation}
where $F_t(E)=(e^{-iEt}-1)/E$, is nothing but the
initial state change induced by its time evolution due to the
exciton-exciton Coulomb scatterings $\xi_{mnij}^\mathrm{dir}$:
We do in particular have $(H-2E_0)B_0^{\dag 2}|v\rangle=
\sum_{mn}\xi_{mn00}^\mathrm{dir}B_m^\dag B_n^\dag|v\rangle$,
where $E_0$ is the energy of the 0 exciton, $(H-E_0)
B_0^\dag|v\rangle=0$ (see eq.\ (6) of ref.\ [6]). In
the absence of these scatterings,
$H|\phi_{00}\rangle$ reduces to
$2E_0|\phi_{00}
\rangle$, so that $(H-\langle H\rangle)|\phi_{00}
\rangle$ reduces to zero, as well as $|\tilde{\psi}_t\rangle$.
The state change $|\tilde{\psi}_t\rangle$ in fact vanishes
linearly with the exciton-exciton Coulomb scatterings, as
$H|\phi_{00}\rangle$ is linear in Coulomb interaction.

From $|\psi_t\rangle$, we can get the initial state
lifetime through
\begin{equation}
e^{-t/\tau_0}= \left|\langle\psi_{t=0}|\psi_t\rangle
\right|^2=1+[\langle\phi_{00}|\tilde{\psi}_t\rangle+c.c]
+\left|\langle\phi_{00}|\tilde{\psi}_t\rangle\right|^2\ .
\end{equation}
Although the dominant term of $|\tilde{\psi}_t\rangle$ is
linear in exciton-exciton Coulomb scatterings, the dominant
term of its scalar product with $|\phi_{00}\rangle$ is
quadratic only. Indeed, if we replace $\langle\phi_{00}|F_t
(H-\langle H\rangle)$ by its zero order contribution, namely
$(-it)\langle\phi_{00}|$, we see, from eq.\ (14), that
$\langle\phi_{00}|\tilde{\psi}_t\rangle$ cancels. We then note
that, as $\langle\psi_t|\psi_t\rangle=1$, the real part of 
$\langle\phi_{00}|\tilde{\psi}_t\rangle$ is just
$(-1/2)\langle\tilde{\psi}_t|\tilde{\psi}_t\rangle$, which is
indeed quadratic in scatterings. So that, to second order in
Coulomb interaction, the lifetime is simply given by
\begin{equation}
\frac{t}{\tau_0}\simeq\langle\tilde{\psi}_t|\tilde{\psi}_t
\rangle\ ,
\end{equation}
since the last term of eq.\ (15) is fourth order.

If we now inject the closure relation between exact
excitons given in eq.\ (12) into this equation (16), we find
\begin{equation}
\frac{t}{\tau_0}\simeq \frac{1}{4}\left[(2-2\lambda_{0000})
\left|\langle\phi_{00}|\tilde{\psi}_t\rangle\right|^2+\sum_{i\neq
0} (2-2\lambda_{iiii})\frac{t}{T_{ii}}+\sum_{i\neq
j}(1-2\lambda_{ijij})\frac{t}{T_{ij}}\right]\ ,
\end{equation}
where the $T_{ij}^{-1}$'s, defined as
$t/T_{ij}=\left|\langle\phi_{ij}
|\tilde{\psi}_t\rangle\right|^2$, are the scattering rates
towards other exciton states induced by the time evolution of 
$|\psi_{t=0}\rangle$. Note that this definition, instead of
$\left|\langle\phi_{ij}|\psi_t\rangle\right|^2$, insures 
these scattering rates to be really linked to the state change
induced by the initial state time evolution, even if the
final states are not orthogonal to the initial state.

As physically expected, and possibly checked from microscopic
calculations [10], these scattering rates contain an energy
conservation which imposes the $(i,j)$ states to be close in
energy to $(0,0)$; so that the states reached by the time
evolution of
$|\phi_{00}\rangle$, with $0=(\nu_0,\v 0)$, must have
the same relative motion index $\nu_0$. Due to momentum
conservation in the scattering processes, these states
in fact are
$i=(\nu_0,\v q)$ and $j=(\nu_0,-\v q)$. Consequently,
$T_{ii}^{-1}=0$ for $i\neq 0$. As the first term of eq.\ (17) is
of the order of the last term of eq.\ (15) we have dropped, we
end with
\begin{equation}
\frac{1}{\tau_0}\simeq\frac{1}{2}\sum_{(i,j)couples}\frac{1}
{T_{ij}}\ ,
\end{equation}
in the large sample limit, since we have shown that, for
excitons having a bound relative motion, the $\lambda_{mnij}$'s
are of the order of the exciton volume divided by the sample
volume, making these
$\lambda$'s negligible in front of 1.

For boson-excitons, the calculation is exactly the same, except
that all the $\lambda_{mnij}$'s are equal to zero, while the
1/4 prefactor of the closure relation (12) is replaced by 1/2.
This change leads to drop the 1/2 in front of the sum in eq.\
(18). This shows in a quite direct way that the relations
between the lifetime and the sum of scattering rates of exact
and boson excitons have to differ by a factor 1/2, in agreement
with the microscopic calculations of $\tau_0$ and the
$T_{ij}$'s we have recently done [10].

In the case of $N$ excitons, the closure relation for exact
excitons contains an additional prefactor $(1/N!)$, instead of
$(1/2!)$ as for $N=2$. It is however clear that this
$(1/N!)$ cannot barely appear in front of the sum of scattering
rates of $N$ exact excitons, otherwise the lifetime of these $N$
excitons would tend to zero in the large
$N$ limit, which is physically unreasonable. And indeed, our
microscopic calculation of $\tau_0$ and the $T_{ij}$'s shows
that the same factor (1/2) exists between $\tau_0^{-1}$ and the
sum of
$T_{ij}^{-1}$'s, as for $N=2$. The proof that these
various factors
$N$ do ultimately disappear, which is not at all trivial, is
beyond the scope of this paper. It is somehow related to
the bosonic enhancement factors we recently found for excitons
embedded in a sea of excitons [19].

\emph{As a conclusion}, this letter allows to clearly show that
the unexpected factor 1/2 we recently found between the lifetime
and the sum of scattering rates of exact excitons, physically
comes from the composite nature of these excitons: They are
deeply made of two fermions and there is no way to get rid of
this fact. This composite nature makes the exciton state set
overcomplete, with a closure relation different from the one of
elementary particles, so that all sum rules deduced from
it have to appear differently. Similar results are \emph{a
priori} expected for composite bosons in other fields than
semiconductor physics. Our letter neatly shows that these
composite bosons should not be reduced to true bosons with an
interaction dressed by exchange, as commonly done: They actually
form a new class of quantum particles, the ``proteons'', their
many-body effects having to be handled
through the new theory for composite bosons we have
recently developed, if we want to fully trust them.

\vspace{0.5cm}

\hbox to \hsize {\hfill REFERENCES
\hfill}

\noindent
(1) D. Hulin, A. Mysyrowicz and C. Benoit \`{a} la Guillaume,
\textit{Phys.\ Rev.\ Lett.\ }\textbf{45}, 1971 (1980).

\noindent
(2) D.W. Snoke, J.P. Wolfe and A. Mysyrowicz, \textit{Phys.\
Rev.\ Lett.\ }\textbf{64}, 2543 (1990).

\noindent
(3) L.V. Butov, A.C. Gossard and D.S. Chemla, \textit{Nature}
\textbf{418}, 751 (2002).

\noindent 
(4) M. Combescot, O. Betbeder-Matibet, K. Cho and H. Ajiki,
\textit{Cond-mat}/0311387.

\noindent
(5) M. Combescot and C. Tanguy, \textit{Europhys.\ Lett}.\ 
\textbf{55}, 390 (2001).

\noindent
(6) M. Combescot and O. Betbeder-Matibet, \textit{Europhys.\
Lett.\ }\textbf{58}, 87 (2002).

\noindent
(7) M. Combescot and O. Betbeder-Matibet, \textit{Europhys.\
Lett.\ }\textbf{59}, 579 (2002).

\noindent
(8) O. Betbeder-Matibet and M. Combescot, \textit{Eur.\ Phys.\
J.\ B} \textbf{27}, 505 (2002).

\noindent
(9) See for example H. Haug and S. Schmitt-Rink, \textit{Prog.\
Quantum Electron.\ }\textbf{9}, 3 (1984).

\noindent
(10) M. Combescot and O. Betbeder-Matibet,
\textit{Cond-mat}/0402087, to be published in \textit{Phys.\
Rev.\ Lett.\ }(2004).

\noindent
(11) M. Greiner, C.A. Regal and D.S. Jin, \textit{Nature}
\textbf{426}, 537 (2003).

\noindent
(12) S. Jochim \textit{et al.,} \textit{Science} \textbf{302},
2102 (2003).

\noindent
(13) M.W. Zwierlein \textit{et al.,} \textit{Phys.\ Rev.\ Lett.\
}\textbf{91}, 250401 (2003).

\noindent
(14) T. Bourdel \textit{et al.,} \textit{Cond-mat}/0403091.

\noindent
(15) M.H. Anderson \textit{et al.,} \textit{Science}
\textbf{269}, 198 (1995).

\noindent
(16) K.B. Davis \textit{et al.,} \textit{Phys.\ Rev.\ Lett.\
}\textbf{75}, 3969 (1995).

\noindent
(17) M. Combescot, X. Leyronas and C. Tanguy, \textit{Eur.\
Phys.\ J.\ B} \textbf{31}, 17 (2003).

\noindent
(18) O. Betbeder-Matibet and M. Combescot, \textit{Eur.\ Phys.\
J.\ B} \textbf{31}, 517 (2003)

\noindent
(19) M. Combescot and O. Betbeder-Matibet,
\textit{Cond-mat}/0403073.

\end{document}